\documentclass[aps,prm,superscriptaddress,reprint,showpacs,floatfix,pdftex]{revtex4-2}
\usepackage[whole]{bxcjkjatype}
\usepackage{graphicx}
\usepackage{comment}
\usepackage{tabularx}
\usepackage{dcolumn}
\usepackage{bm}
\usepackage{amsmath}
\usepackage{amssymb}
\usepackage{txfonts}
\usepackage{soul}
\usepackage{xcolor}
\usepackage{multirow}
\usepackage{ulem}
\usepackage{subfigure}
\usepackage{physics}
\usepackage{braket}
\usepackage{mathrsfs}
\usepackage{empheq}

\usepackage{xurl}


\begin{document}
\title{
  Application and Performance Assessment of Annealing Methods for
  Electrostatic-Energy-Based Configuration Search in Mixed Crystals
}
\date{\today}
\author{Tack Saquai }
  \affiliation{School of Information Science, JAIST,
  Asahidai 1-1, Nomi, Ishikawa 923-1292, Japan}
\author{Kenta Hongo}
\affiliation{
  Center for Advanced Scientific Computing,
  JAIST, Asahidai 1-1, Nomi, Ishikawa 923-1292, Japan}
\author{Ryo Maezono}
\affiliation{
  Graduate Major in Materials and Information Sciences,
  School of Materials and Chemical Technology,
  Institute of Science Tokyo
  2-12-1-S6-22 Ookayama, Meguro-ku, Tokyo 152-8550, Japan.
}
\author{Tom Ichibha}
\email[]{ichibha@gmail.com}
\affiliation{School of Information Science, JAIST, Asahidai 1-1, Nomi, Ishikawa 923-1292, Japan}
\begin{abstract}
  In first-principles design of solid solutions and disordered materials,
  exhaustive evaluation of all substitutional configurations is often
  impractical because the number of possible occupations grows exponentially.
  We developed a framework for pre-screening mixed-crystal configurations
  using electrostatic energy evaluated by Ewald summation as the objective
  function. Substitutional occupations were represented by binary variables,
  and the Ewald energy was mapped onto an Ising-type Hamiltonian, enabling
  combinatorial optimization by simulated annealing (SA) and quantum
  annealing (QA). Performance was evaluated against exhaustive search in
  terms of computational speed-up and recovery of the lowest-energy
  structures. 
  For the small-scale CaYAlO$_4$, SA and QA achieved speed-ups of 
  about 30 and over 100 times, respectively, while recovering all lowest-energy 
  configurations. For the medium-scale $\beta$-KSbF$_4$ and larger Ba-doped 
  SiAlON systems, SA achieved speed-ups of roughly 200--300 times while stably 
  identifying the lowest-energy structures.
  In contrast, QA showed limited speed-up for the
  medium-scale $\beta$-KSbF$_4$ system and missed some low-energy
  configurations, likely due to chain breaks. These results indicate that
  SA is currently a robust approach for rapid electrostatic-energy-based
  pre-screening of mixed-crystal configurations, including larger systems.
  The framework can be implemented with publicly available libraries and
  integrated with first-principles calculations to accelerate
  candidate-structure generation and property prediction.
\end{abstract}
\maketitle

\section{Introduction}
In the materials design of solid solutions and disordered systems,
properties are often tuned to a desired range by introducing dopants
such as cations, anions, or vacancies~\cite{2018_HK_KP}. When such
systems are treated using first-principles calculations, one must
confront the combinatorial number of possible elemental arrangements
at the substitutional sites~\cite{2016_KO_SC, 2022_GP_KN}. This number
can increase exponentially with the number of substitutional sites
~\cite{2022_GP_KN}. For example, in the structural model of
Zn$_{0.5}$Fe$_{2}$Cu$_{0.5}$O$_{4}$ reported in COD 9012442
~\cite{2009Saulius_Armel}, as many as 127,400 symmetry-distinct configurations
are possible within the unit cell specified therein.
In magnetic systems, the number can become even larger
because substitutional sites occupied by the same ionic species may
still be classified as inequivalent depending on their spin orientations,
reaching as many as 1,229,107,765,600 possible configurations
~\cite{2021_KU_RM}.

\vspace{2mm}
These configurations generally have different energy values~\cite{2022Jin_Li}.
However, in practical first-principles simulations, the configurations
that should be considered are usually limited to a subset of patterns
with sufficiently low and physically realizable energies~\cite{2024_KI_TY}.
The most straightforward procedure for identifying such configurations would be
to perform structural relaxation using first-principles calculations
for all possible substitutional-site occupation patterns, compare
their energies, identify the most stable structures, and then proceed
to property calculations. However, screening hundreds of millions or
even trillions of structures by evaluating their energies one by one
using first-principles calculations is not practically feasible.

\vspace{2mm}
This naturally leads to a strategy in which configurations
are screened not by using accurate energy values, but by
using an alternative descriptor that can be evaluated much
more rapidly. One possible descriptor for this purpose is
the electrostatic energy~\cite{2023_SJ_YT}. The total energy
of a material system is determined by the distribution of
many interacting electrons on top of the classical electrostatic
interaction energy arising from the ionic arrangement~\cite{2018_CP}.
First-principles calculations are designed to treat the delicate
balance associated with the former electronic contribution
~\cite{2023_MR, 2018_CP}.
In ionic crystals, however, the relative
stability of possible substitutional-site configurations can
often be governed primarily by the latter electrostatic
interaction~\cite{2023_SJ_YT, 1987_MT_MS}.
Indeed, for structural models of solid solutions with
different ionic configurations, a strong correlation
between the electrostatic energy and the energy obtained
from first-principles calculations has been reported
~\cite{2023_SJ_YT, 2024TB_MHE, 2025Koster_Kaghazchi, 2022Lee_KeeSun}.
This electrostatic contribution is commonly discussed in terms of the Ewald energy.
The Ewald energy can be evaluated much faster than
a first-principles calculation. For example, in the present
system, its evaluation is approximately 43,000 times faster.
Extrapolating this speed-up to all 127,400 configurations,
the screening would require approximately 146,000 h using
first-principles calculations, whereas it would take only
about 3 h when based solely on Ewald-energy evaluations.

\vspace{2mm}
Software packages that can rapidly estimate Ewald energies
are widely available~\cite{2013Ong_Ceder, 2003Gale_Rohl}, and one might therefore expect that screening
can simply be performed using such tools. However, even with this
reduced evaluation cost, the computational time becomes prohibitive
when realistic mixed-crystal materials are considered. For example,
in the case of (Nd$_{0.7}$Ce$_{0.225}$La$_{0.075}$)$_2$Fe$_{14}$B,
even if the evaluation time is only about 0.1 s per configuration,
the total number of possible configurations reaches 1,229,107,765,600,
resulting in an unrealistic computational time of approximately
34,000,000~h, or about 3,880 years~\cite{2021_KU_RM}. Thus, an
exhaustive search over all possible substitutional-site occupation
patterns is impractical.
Since our ultimate interest lies in substitutional structures
with low energies, it is desirable to develop a method that
efficiently eliminates high-energy substitutional configurations
and selectively extracts low-energy ones.

\vspace{2mm}
As a method for such selective extraction, we considered the application
of annealing methods~\cite{2024TB_MHE}. To implement an annealing method,
it is necessary to formulate a model in which the objective function is
expressed in terms of binary variables. This enables the optimization
problem to be mapped onto the problem of searching for the most stable
spin configuration.
As described in \S\ref{sec.method}, the occupation of substitutional
sites can be represented by 0/1 binary variables, and the Ewald
energy can be expressed as a function of these binary variables,
which we refer to as the annealing model. By using this value
as the objective function, the problem can be formulated as a combinatorial
optimization problem in which one searches for a bit string that gives a
lower Ewald energy.
In general, combinatorial optimization problems are algorithmically
challenging because they are often accompanied by multimodality
~\cite{1983_SK_MV}.
Heuristic search methods for addressing such problems have
therefore been actively studied. Examples include genetic algorithms
~\cite{2021_PS_RM, 2020_TA_MB}, particle swarm optimization
~\cite{2021_TO_FR, 1995_RE_JK}, and Bayesian optimization
~\cite{2020_TY_KH, 2016_BS_NF}.
For an objective function expressed in terms of binary variables,
annealing techniques can be applied by mapping the problem onto a spin
Hamiltonian. More specifically, the problem can be recast as the search
for the spin configuration that realizes the lowest energy. This problem
can be solved using methods such as quantum annealing
~\cite{2021_KU_RM, 2024_KI_TY} and simulated annealing
~\cite{1983_SK_MV}.

\vspace{2mm}
As described in \S\ref{sec.results}, annealing-based extraction of
low-energy substitutional structures does not necessarily guarantee
complete coverage of all low-energy candidates. Although a previous
study demonstrated that quantum annealing can mitigate this issue
~\cite{2024TB_MHE}, the extent of the speed-up achievable by annealing
methods and their ability to recover all relevant low-energy structures
without omission have not been quantitatively clarified. The present
study therefore aims to evaluate these two aspects systematically.

\vspace{2mm}
Previous studies have explored Ewald-energy-based searches
over configuration spaces far larger than that considered
in the present study, using high-throughput enumeration,
swap-based local optimization, and various combinatorial
optimization methods~\cite{2023_SJ_YT, 2025Koster_Kaghazchi}.
However, these studies did not examine how comprehensively
low-energy configurations, including the global minimum,
were covered. In contrast, the present study deliberately
restricts the system size to one for which exhaustive
enumeration is feasible. By obtaining the complete
distribution over the entire configuration space,
we evaluate how efficiently and comprehensively
the annealing methods can identify low-energy configurations.

\vspace{2mm}
Based on the above considerations, the objective of the present study is
defined as follows. We consider Ewald-energy-based screening of
a large number of substitutional structural models for mixed-crystal
systems of inorganic compounds. The Ewald energy is expressed as an
annealing model, and the problem is formulated as a combinatorial
optimization problem in which the binary variables that minimize the
energy are searched for.
This combinatorial optimization problem is solved primarily
using simulated annealing~\cite{1983_SK_MV}, while quantum annealing
~\cite{1998_TK_HN} is also examined as a complementary approach.
By comparing the results with those obtained
by exhaustive search, we quantitatively evaluate the speed-up achieved by
annealing methods and their ability to identify low-energy structures
without omission.

\vspace{2mm}
In this study, we selected three compounds according to the number of
possible substitutional-site configurations, which corresponds to the
difficulty of the combinatorial optimization problem: CaYAlO$_4$, which
has a relatively small number of configurations; $\beta$-KSbF$_4$, which
has a more realistic problem size; and Ba-doped SiAlON, which represents
an even larger-scale system.
In selecting these compounds, we chose systems in which the ions
substituted in the mixed crystal have valences different from those of
the host ions, so that the Ewald energy, used as the objective
function, varies among different configurations.

\vspace{2mm}
For the small-scale system CaYAlO$_4$, both simulated annealing and
quantum annealing were able to efficiently extract only mixed-crystal
configurations with low Ewald energies. Compared with the
conventional exhaustive search, simulated annealing achieved a speed-up
of nearly 30 times and identified the lowest-energy structures without
missing any candidates. For a problem of this size, quantum annealing
performed even faster, achieving a speed-up of more than 100 times while
also identifying the lowest-energy structures without omission.
For $\beta$-KSbF$_4$, which represents a realistic level of difficulty,
simulated annealing achieved a speed-up of more than 200 times depending
on the cooling schedule. In contrast, quantum annealing did not exhibit
sufficient performance.
This was attributed mainly to constraints in the
implementation of physical qubits, such as chain breaks. This finding
provides new insight that was not discussed in the previous study~\cite{2024TB_MHE}.
Simulated annealing also achieved a speed-up of more than 300 times for
Ba-doped SiAlON, an even larger and more challenging system. These results
indicate that mixed-crystal configuration search can potentially be
performed extremely rapidly when formulated as an annealing problem.
At the same time, they also show that, for quantum annealing to fully
demonstrate its performance in this direction, constraints arising from
the current implementation of physical qubits remain an important issue.

\vspace{2mm}
The remainder of this paper is organized as follows.
In \S\ref{sec.method}, we describe the formulation and implementation
of the Ewald-energy evaluation of
substitutional-site configurations
as a combinatorial optimization problem by mapping it onto an Ising-type
Hamiltonian and applying annealing methods. We also describe the target
systems used for validation.
In \S\ref{sec.results}, we evaluate the performance of simulated
annealing and quantum annealing for systems of different sizes by
comparing their search results with those obtained by exhaustive search.
We also discuss, from the viewpoint of chain breaks, the factors that
cause quantum annealing to miss optimal solutions in medium-scale and
larger systems.
Finally, \S\ref{sec.conclusion} summarizes the conclusions of this study.

\section{method}
\label{sec.method}
We consider the Ewald energy $E$ of a periodic system with site
substitutions in a solid solution. Let $q_i$ denote the charge of a site
that can be substituted, and $Q_k$ denote the charge of a site that is
known in advance not to be substituted, where $i$ and $k$ are site
indices. The Ewald energy $E$ can be expressed as a superposition of
pairwise interactions:
\begin{equation}
  E = \sum_{j}^{}\sum_{i\leq j}^{}A_{ij}q_{i}q_{j}+\sum_{k}^{}
  \sum_{l\leq k}^{}B_{kl}Q_{k}Q_{l}+\sum_{k}^{}\sum_{i}^{}C_{ki}Q_{k}q_{i}
  \label{eq.ewald}
\end{equation}
where $A_{ij}$, $B_{kl}$, and $C_{ki}$ are interaction coefficients
determined solely by the geometric arrangement of the sites.
The procedure for explicitly determining these parameters
for a given disordered crystal structure is described in
Sec.~S1 of the Supplementary Information.

\vspace{2mm}
Whether the site charge $q_i$ is actually substituted or not can be
represented by a binary variable $\sigma_{i} = 0$ or 1. Let
$q_{i}^{\left(0\right)}$, and $q_{i}^{\left(1\right)}$ denote
the charges before and after substitution, respectively.
Then, $q_i$ can be written as
\begin{equation}
  q_{i}=\left(1-\sigma_{i}\right)\cdot q_{i}^{(0)}+\sigma_{i}\cdot q_{i}^{(1)}\,\,\,\,\left(\sigma_{i}=\mathrm{0}\,\mathrm{or}\,\mathrm{1}\right).
  \label{eq.charge}
\end{equation}
Substituting Eq.~\eqref{eq.charge} into
Eq.~\eqref{eq.ewald}, the energy
can be rearranged into the form of an
Ising-type Hamiltonian as
\begin{equation}
  E=
  \sum_{i\leq j}^{}J_{ij}\sigma_{i}\sigma_{j}+
  \sum_{i}^{}h_{i}\sigma_{i}+\mathrm{const.}
  \label{eq.ising}
\end{equation}
The original electrostatic interaction coefficients
$A_{ij}$, $B_{kl}$, and $C_{ki}$ can be determined once the crystal
structure is specified. Therefore, the model parameters $J_{ij}$ and
$h_i$ can be obtained from these coefficients.

\vspace{2mm}
Thus, each possible substitutional-site occupation state can be
represented as a spin configuration
$\boldsymbol{\sigma} = (\sigma_1, \sigma_2, \dots, \sigma_N)$.
The problem can therefore be recast as an annealing problem in which
the Ising-type Hamiltonian is minimized over all possible spin
configurations $\left\{0,1\right\}^N$.

\vspace{2mm}
To restrict the search space to solid solutions with a prescribed site
substitution ratio, an additional constraint term is added to
Eq.~\eqref{eq.ising}. Let $\alpha$ be an index that distinguishes the
types of substitutional sites, such as elemental species or inequivalent
Wyckoff positions, and let $N_{\alpha}$ denote the total number of
substituted sites corresponding to each substitution ratio. The
Hamiltonian is then given by
\begin{equation}
  H_{\mathrm{ewald}} =
  \sum_{i\leq j}^{}J_{ij}\sigma_{i}\sigma_{j}
  +\sum_{i}^{}h_{i}\sigma_{i}+
  \lambda\sum_{\alpha}^{}\left(\sum_{i\in\alpha}^{}
  \sigma_{i}-N_{\alpha}\right)^2.
  \label{eq.hamiltonian}
\end{equation}
Here, $\lambda$ is a coefficient that controls the strength of the
constraint. Specifying the substitutional composition of a solid solution
corresponds to specifying the set $\left(N_1,N_2,\cdots\right)$.
Using $H_{\mathrm{ewald}}$ as the objective function, we performed a
search for the optimal Ising spin configuration
$\boldsymbol{\sigma} = \left(\sigma_1, \sigma_2, \dots, \sigma_N\right)$
that minimizes this function, using simulated annealing and quantum
annealing.

\vspace{2mm}
As described above, the present formulation represents
the two states before and after substitution by assigning
a single binary variable to each substitutional site.
However, in many practically important solid-solution systems,
such as high-entropy materials and multielement-doped systems,
more than two elemental species may occupy the same Wyckoff position,
which cannot be represented within the present formulation.
In such cases, the formulation can be extended as follows to
obtain a Hamiltonian analogous to that in Eq.~\eqref{eq.hamiltonian}.

\vspace{2mm}
To represent which element $e$ occupies an atomic site $i$, we introduce
a binary variable $\sigma_{i,e}$. The variable takes a value of 1 when
site $i$ is occupied by element $e$, and 0 otherwise. If the charge of
element $e$ is $q^{\left(e\right)}$, the charge $q_i$ at atomic site $i$
can therefore be expressed as
\begin{equation*}
  q_i=\sum_{e}q^{\left(e\right)}\sigma_{i,e}.
\end{equation*}
Since each atomic site $i$ must be occupied by exactly one element,
only one of the variables $\sigma_{i,e}$ associated with a given $i$
can take a value of 1, while all the others must be 0. This constraint
can be imposed by introducing
\begin{equation*}
  \left\{\left(\sum_{e}\sigma_{i, e}\right)-1\right\}^2
\end{equation*}
as a constraint term in the Hamiltonian.
When multiple substitutional sites $i$ are present,
the condition that each site is occupied by exactly
one element can be expressed by the constraint term
\begin{equation*}
  \sum_{i}\left\{\left(\sum_{e}\sigma_{i, e}\right)-1\right\}^2.
\end{equation*}

\vspace{2mm}
A constraint must also be imposed to satisfy the prescribed site
occupancies. For a given number $N_{\alpha,e}$ of atoms of element $e$
occupying Wyckoff position $\alpha$, this constraint can be expressed by
introducing
\begin{equation*}
  \left\{\left(\sum_{i\in\alpha}\sigma_{i, e}\right)-N_{\alpha, e}\right\}^2
\end{equation*}
as a constraint term.
Since this condition must be satisfied for all
Wyckoff positions $\alpha$ and elements $e$,
it can generally be written as
\begin{equation*}
  \sum_{e, \alpha}\left\{\left(\sum_{i\in\alpha}\sigma_{i, e}\right)-N_{\alpha, e}\right\}^2.
\end{equation*}
Because the total number of atomic sites belonging to each Wyckoff
position $\alpha$ is fixed, for a system containing $c$
components, only $c-1$ independent constraints with
respect to the elements are required. As demonstrated above, 
the present formulation can therefore
be extended to more general solid-solution systems, including
multielement-doped systems~\cite{2025Koster_Kaghazchi}.

\vspace{2mm}
As target systems, we considered several compounds according to the
number of possible substitutional-site configurations, which corresponds
to the difficulty of the combinatorial optimization problem. We selected
CaYAlO$_4$~\cite{1992Shannon_Parise} as the small system with the
smallest number of configurations, and $\beta$-KSbF$_4$
~\cite{1999Yamada_Okuda} as the medium system with a more realistic
problem size. For simulated annealing, we also considered Ba-doped
SiAlON~\cite{2004Esmaeilzadeh_Thiaux} as a large and more challenging
system with an even larger number of configurations.
The calculations were performed using a $1\times 2\times 2$ supercell
for CaYAlO$_4$, a $1\times 1\times 2$ supercell for $\beta$-KSbF$_4$,
and a $2\times 2\times 1$ supercell for Ba-doped SiAlON. With this
choice of supercells, all elements of the set $\left\{N_\alpha\right\}$ can be
taken as integers. A comparison of the number of possible substitutional
configurations for these systems is shown in Table~\ref{tab:target_system}.
In these compounds, the substituting ions have valences different from
those of the ions being substituted. Therefore, the Ewald energy,
which serves as the objective function for optimization, varies among
different configurations, making these systems suitable test cases for
optimization.
CaYAlO$_4$ is an important host material for phosphors and lasers
~\cite{2020Xia_Zhou, 2023Liu_Tu}, $\beta$-KSbF$_4$ is an important ionic
conductor~\cite{1999Yamada_Okuda, 2025Hisasue_Miura}, and Ba-doped
SiAlON is an important host material for phosphors
~\cite{2008Duan_Hintzen}.

\vspace{2mm}
For each system considered, the charges assigned to the respective
elements were given as listed in Table~\ref{tab:ionic_charges}.
These charge assignments preserve overall charge neutrality in each
system and are consistent with the common oxidation states or formal
charges of the corresponding elements. Several previous studies have
reported that, with charges assigned in this manner, the electrostatic
energies of different ionic configurations in solid solutions correlate
well with the energies obtained from first-principles calculations
~\cite{2024TB_MHE, 2023_SJ_YT, 2025Koster_Kaghazchi}.
These charge assignments are therefore appropriate for the present
purpose of searching for stable mixed-crystal configurations.

\begin{table}[t]
  \centering
  \caption{Charges assigned to each element in the present study.}
  \label{tab:ionic_charges}
  \begin{tabular}{cc@{\qquad}cc@{\qquad}cc}
    \hline
    Element & Charge & Element & Charge & Element & Charge \\
    \hline
    Ca      & $+2$   & K       & $+1$   & Ba      & $+2$   \\
    Y       & $+3$   & Sb      & $+3$   & Si      & $+4$   \\
    Al      & $+3$   & F       & $-1$   & Al      & $+3$   \\
    O       & $-2$   &         &        & O       & $-2$   \\
            &        &         &        & N       & $-3$   \\
    \hline
  \end{tabular}
\end{table}

\begin{table*}[htbp]
  \centering
  \caption{
    Number of substitutional sites in the simulation cell and the number of
    possible substitutional configurations. `SA[G]' and `SA[L]' denote
    simulated annealing with geometric and linear cooling schedules,
    respectively, and `QA' denotes quantum annealing. The values listed for
    these methods indicate the speed-up factors relative to exhaustive search.
    Because the speed-up depends on the computational conditions of the
    annealing methods, the listed values should be interpreted as
    representative speed-up factors achieved in the present calculations.
    For the largest system, Ba-doped SiAlON, the QA result is not shown
    because the calculation could not be performed, and the SA[L] result is
    omitted because low-energy configurations were missed.
  }
  \label{tab:target_system}
  \begin{tabular}{lccccccc}
    \hline
    Target system                                                     &
    \shortstack{Number of mixed-\\crystal sites}                      &
    \shortstack{Number of possible \\ substitutional configurations } &
    \shortstack{Number of symmetrically\\ inequivalent structures}    &
    \shortstack{Number of \\samples}                                  &
    SA[G]                                                             & SA[L] & QA                                                  \\
    \hline
    CaYAlO$_4$~($1\times 2\times 2$)
                                                                      & 8     & 70        & 10       & 30      & 26.6 & 43.8 & 116  \\
    $\beta$-KSbF$_4$~($1\times 1\times 2$)
                                                                      & 18    & 794       & 52       & 250     & 18.8 & 248  & 1.29 \\
    Ba-doped SiAlON~($2\times 2 \times 1$)
                                                                      & 24    & 127{,}400 & 31{,}976 & 5{,}000 & 286  & --   & --   \\
    \hline
  \end{tabular}
\end{table*}

\vspace{2mm}
To implement annealing based on Eq.~\eqref{eq.hamiltonian}, we used two
methods in this study: simulated annealing (SA) and quantum annealing
(QA). In both methods, we perform \(M\) independent annealing runs and obtain
one spin configuration as a sample from each run.
These samples are then
individually annealed toward solutions of Eq.~\eqref{eq.hamiltonian},
and the resulting spin configurations are observed.
The way in which this relaxation process is implemented differs between
SA and QA. The observed set of $M$ Ising spin configurations,
$\left\{\boldsymbol{{\sigma}}^{\left(j\right)}\right\}_{j=1}^{M}$,
corresponds to $M$ substitutional-site configurations that realize
low-energy states. In this way, substitutional-site configurations can be obtained.

\vspace{2mm}
For simulated annealing, we used the \texttt{dwave-samplers} library.
The number of samples $M$ was set to 30 for CaYAlO$_4$, 250 for
$\beta$-KSbF$_4$, and 5{,}000 for Ba-doped SiAlON
~(Table~\ref{tab:target_system}).
In simulated annealing,
the temperature associated with the Boltzmann
factor for Eq.~\eqref{eq.hamiltonian} is gradually decreased from a high
temperature to a low temperature.
At each temperature step, the spin
configuration is relaxed toward equilibrium according to the Boltzmann
factor at that temperature. This procedure is repeated while updating
the temperature step by step.
At finite temperatures, changes that increase the energy can also be
accepted probabilistically. In addition, changes in the spin
configuration may temporarily alter the composition
at the cost of incurring the penalty $\lambda$. These features
allow individual samples to escape from local minima and potentially
explore configurations with lower energies in an energy landscape
containing multiple local minima.
The rule used to update the temperature is
called the cooling schedule.
For the cooling schedule in this study, we applied and compared
geometric cooling,
\begin{align}
  \beta_{k+1}
   & =
  \left(\cfrac{\beta_{N}}{\beta_{1}}\right)
  ^{\frac{1}{N-1}}\beta_{k},
  \quad (1 \leq k \leq N-1)
  \label{eq.geometric}
\end{align}
and linear cooling,
\begin{align}
  \beta_{k+1}
   & = \beta_{k}
  + \cfrac{\beta_{N} - \beta_{1}}{N-1},
  \quad (1 \leq k \leq N-1)
  \label{eq.linear}
\end{align}
where $\beta_k$ is the inverse temperature at the $k$-th step.
These equations represent the cooling update schemes over $N$ steps.
In the \texttt{dwave-samplers} library, the initial value $\beta_{1}$
and the final value $\beta_{N}$ used for the update are given by
\begin{equation*}
  \beta_{1} = \cfrac{\log_e 2}{\Delta E_{\mathrm{max}}}
  \quad ,\quad
  \beta_{N} = \cfrac{\log_e 100}{\Delta E_{\mathrm{min}}}.
\end{equation*}
Here, $\Delta E_{\max}$ and $\Delta E_{\min}$ are, respectively, the
approximate upper and lower bounds on the energy increase
$\Delta E$ that occurs when a single spin is flipped from an arbitrary
spin configuration $\boldsymbol{\sigma}$.
The upper and lower bounds of the energy increase,
$\Delta E_{\mathrm{max}}$ and $\Delta E_{\mathrm{min}}$, respectively,
are determined from the Hamiltonian in Eq.~\eqref{eq.hamiltonian},
including the constraint terms.
At the initial inverse temperature $\beta_{1}$, even a spin flip
associated with an energy increase of $\Delta E_{\mathrm{max}}$ is
accepted with a probability of 50\%.
Since $\Delta E_{\mathrm{max}}$ is the upper bound of the energy change
in Eq.~\eqref{eq.hamiltonian}, most energy-increasing changes are smaller
than this value and are therefore accepted with probabilities greater
than 50\%. At the final stage, where the inverse temperature reaches
$\beta_{N}$, even the minimum energy increase $\Delta E_{\mathrm{min}}$
is accepted with a probability of only 1\%, thereby promoting convergence
toward the optimal solution. Since most energy increases are larger than
$\Delta E_{\mathrm{min}}$, their acceptance probabilities are even
lower than 1\%.
The total number of steps used for cooling, $N$,
is referred to as the number of sweeps. In this study, we basically
adopted geometric cooling with $N=1{,}000$ sweeps. This value is the
default setting in the \texttt{dwave-samplers} library.

\vspace{2mm}
For the hardware implementation of quantum annealing, we used D-Wave's
\texttt{Advantage system 4.1} QPU~\cite{2020McGeoch_Farre}. The
Hamiltonian was embedded onto the quantum computer using the
\texttt{AutoEmbeddingComposite} in the \texttt{Ocean SDK}~\cite{2024DWave}.
In quantum annealing, the search for the optimal solution is performed
by taking an equal superposition of all possible states, namely the
Hadamard state, as the initial state and evolving the system toward the
ground state of the given Hamiltonian~\cite{1998_TK_HN}. This operation
is realized by varying the Hamiltonian with time. Specifically, the
time-dependent Hamiltonian is taken as the linear combination of the
transverse-field term $H_{\rm trans}$, whose ground state is the
Hadamard state, and the problem Hamiltonian $H_{\rm ewald}$:
\begin{equation}
  H\left(t\right)
  =
  A\left(\cfrac{t}{\tau}\right)\cdot H_{\rm{trans}}
  + B\left(\cfrac{t}{\tau}\right)\cdot H_{\rm{ewald}}
  , \quad  \left( 0 \leq t \leq \tau \right).
  \label{eq.timeDep}
\end{equation}
The Hamiltonian is then varied with time $t$.
The functions $A(s)$ and $B(s)$ are determined by the implementation of
the quantum computer. They satisfy $A\left(s\right) \gg B\left(s\right)$
at $s=0$ and $A\left(s\right) \ll B\left(s\right)$ at $s=1$.
In other words, as $s$ increases, the contribution of the
transverse-field term approaches zero. Here, $\tau$
is called the annealing time and determines the rate at which the
contribution of the transverse-field term is reduced. In this study,
$\tau$ was varied in the range from 20~$\mu$s to 2{,}000~$\mu$s.
According to Eq.~\eqref{eq.timeDep}, the effective coupling constants
between Ising spins are varied over the time interval
$t=\left[0,\tau\right]$, and the relaxed spin configurations are obtained.
The number of samples $M$ was set to 30 for CaYAlO$_4$ and 250 for
$\beta$-KSbF$_4$.

\vspace{2mm}
In quantum annealing, the ideal implementation is that each Ising
variable is represented by a single qubit. However, owing to technical
constraints, a workaround is used in which a single Ising variable is
represented by multiple physical qubits~\cite{2015Vinci_Lidar}. A group
of physical qubits corresponding to a single Ising variable is called a
chain.
Ideally, all physical qubits forming a chain are expected to take the
same state. In practice, however, an undesired situation often occurs in
which the physical qubits within a chain do not take the same state.
This situation is referred to as a chain break. Chain breaks will be
discussed later in this paper.
When a chain break occurs, we determine the corresponding Ising variable
by taking a majority vote over the states of the physical qubits in the
chain. The physical qubits within a chain are coupled ferromagnetically
so that they tend to take the same state. The strength of this coupling
is a hyperparameter called the chain strength, denoted by $\mu$.
Increasing the chain strength makes chain breaks less likely to occur.
On the other hand, it is known that an excessively large chain strength
can degrade the performance in finding the optimal 
solution~\cite{2022Grant_Humble}. Therefore, the chain strength 
should not simply be made as large as possible; an appropriate 
value must be selected.

\vspace{2mm}
In addition to the chain strength, another tunable hyperparameter is the
coefficient $\lambda$ of the constraint term in Eq.~\eqref{eq.hamiltonian}.
Although these hyperparameters should be appropriately determined
depending on the problem, no general method for determining their optimal
values has been established, and this remains an active area of research
~\cite{2023Roch_Feld, 2022Willsch_Michielsen, 2008Pedamallu_Ozdamar}.
The value of $\lambda$ was determined by monitoring whether the
constraint represented by the corresponding constraint term was satisfied.
When $\lambda$ is sufficiently large, the constraint is satisfied.
We then gradually reduced this value and identified the smallest value
for which the constraint was satisfied as well as possible.
For simulated annealing, we used $\lambda^{\left({\rm SA}\right)}=50$~eV
for all systems. For quantum annealing, we used
$\lambda^{\left({\rm QA}\right)}=150$~eV for the small-scale system
CaYAlO$_4$ and $\lambda^{\left({\rm QA}\right)}=700$~eV for the
medium-scale system $\beta$-KSbF$_4$.
The chain strength $\mu$ was determined by searching for a value that
made the chain break fraction $B$ as small as possible. We used
$\mu=250$ for the small-scale system CaYAlO$_4$ and $\mu=1{,}400$ for
the medium-scale system $\beta$-KSbF$_4$.

\vspace{2mm}
For calibration of the results, we also reconstructed the set of charges
$\left\{q_i\right\}$ from each spin configuration
$\boldsymbol{\sigma} =
  \left(\sigma_1, \sigma_2, \dots, \sigma_N\right)$
and evaluated the Ewald energy using Eq.~\eqref{eq.ewald}.
This enables a direct comparison with the Ewald-energy evaluation
performed by the conventional exhaustive search, in terms of both the
calculated energy values and the computational cost. For the calculated
energy values, the difference between the two approaches was confirmed
to be on the order of $10^{-9}$~eV or less per simulation cell.
Because Eq.~\eqref{eq.hamiltonian} accurately reproduces the conventional
electrostatic energy, its optimization can be used to search for
low-energy ionic configurations corresponding to a given set of charges.
This correspondence does not rely on any specific choice of charge
values. Therefore, the corresponding Hamiltonian can be constructed in
the same manner for different charge assignments, allowing the present
method to be applied without modification. In this sense, the method is
robust with respect to the choice of charges for the target system.

\vspace{2mm}
The hardware specifications of the computer used for simulated annealing
and exhaustive search, together with the major software versions used in
this study, are summarized in Supplementary Table~S1.
In the exhaustive search, the Ewald energy was evaluated sequentially
for all substitutional configurations, including symmetrically equivalent ones.

\section{Results and discussion}
\label{sec.results}
For both the small- and medium-scale systems, simulated annealing showed
good performance improvement for all systems. In contrast, quantum
annealing achieved sufficient performance improvement only for the
small-scale system, owing to limitations in the current hardware
implementation, and its performance improvement was limited for the
medium-scale system~(Table~\ref{tab:target_system}).
In this section, we first present the results of simulated annealing
~(\S\ref{sec:results.sa}), followed by the results obtained 
when quantum annealing was applied to the same problem~(\S\ref{sec:results.qa}).
Optimization is expected to proceed more easily when the number of binary
variables describing the system is smaller, that is, when the number of
substitutional sites in the crystal is smaller. Therefore, we begin with
the small-scale system, then increase the system size, and compare the
results.
We then examine the case of a larger-scale system
~(\S\ref{sec:results.large}). The effect of the cooling schedule in SA
is discussed in Sec.~S2 of the Supplementary Information. 
Finally, we discuss incomplete sampling in QA
and its possible relationship with chain breaks
~(\S\ref{sec:results.omission}).

\subsection{Simulated annealing for small- and medium-scale systems}
\label{sec:results.sa}
\begin{figure*}[t]
  \centering
  \subfigure[Small-scale system CaYAlO$_4$]{%
    \includegraphics[width=0.3\linewidth]{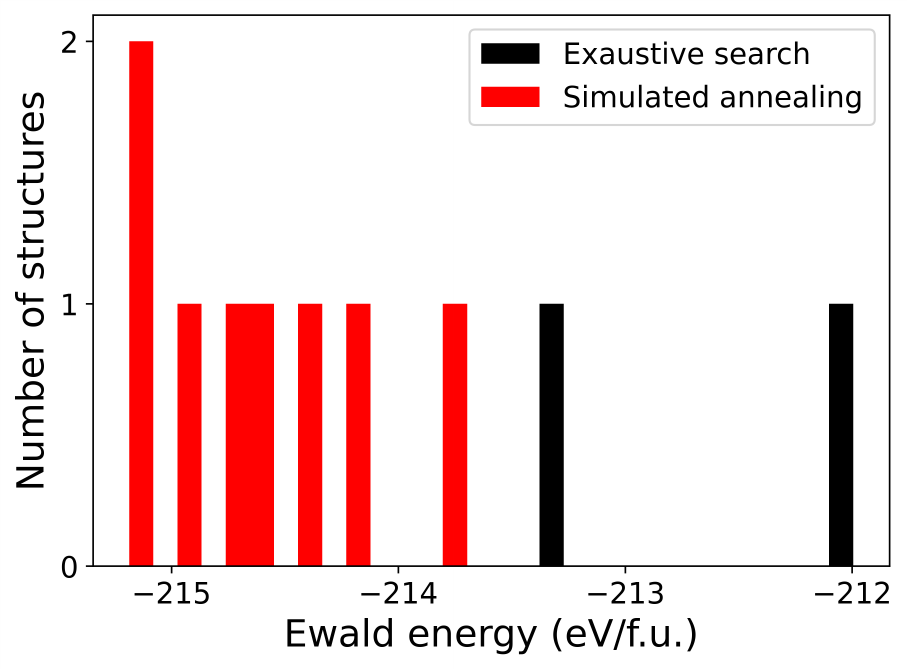}%
    \label{fig:CaYAlO4_SA_unique_hist}}
  \subfigure[Medium-scale system $\beta$-KSbF$_4$]{%
    \includegraphics[width=0.3\linewidth]
    {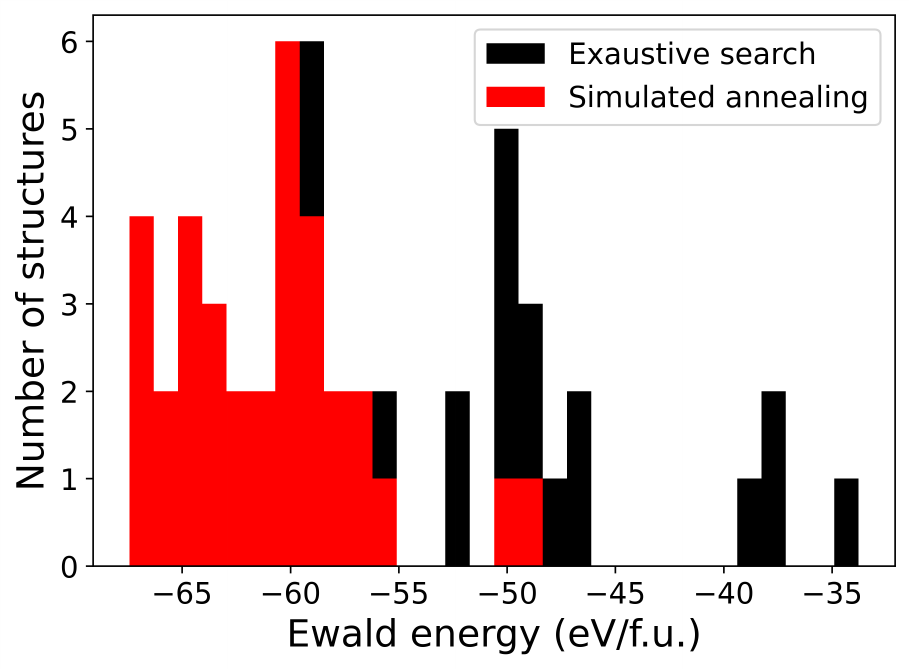}%
    \label{fig:KSbF4_SA_unique_hist}}
  \caption{
    Distributions of Ewald energies for all symmetrically
    inequivalent mixed-crystal structures obtained by exhaustive search
    (black) and for the structures obtained by simulated annealing
    (red) in (a) the small-scale system CaYAlO$_4$ and
    (b) the medium-scale system $\beta$-KSbF$_4$.
    The Ewald energy is given in eV per formula unit.
    The red distributions cover the low-energy regions, indicating that
    low-energy substitutional-site configurations were captured without
    omission.
  }
  \label{fig:SA_unique}
\end{figure*}

Figure~\ref{fig:SA_unique} shows the results obtained by applying
simulated annealing to the small-scale system [panel (a)] and the
medium-scale system [panel (b)]. Here, the default geometric cooling
schedule was used. The results obtained with linear cooling are presented
in Sec.~S2 of the Supplementary Information.
In this figure, the horizontal axis represents the energy values of the
sampled site-substitution patterns, and the vertical axis shows the
number of configurations in each energy range as a histogram.
Because the number of possible atomic configurations is
limited, the histogram appears discrete. In constructing the histogram,
we also took into account the fact that configurations that are different
as spin configurations can be identical as crystal structures owing to
spatial symmetry. Therefore, the vertical axis represents the accumulated
number of mutually inequivalent crystal structures belonging to the same
energy range.
If the histogram obtained by exhaustive search (black) is not fully
covered by that obtained by the annealing method (red), this indicates
that some crystal structures were missed by the annealing method. Since
the purpose of pre-screening is to extract the crystal structures with
the lowest energy, the key point is whether the red histogram fully
covers the lowest-energy bin.

\vspace{2mm}
For both the small-scale system [panel (a)] and the medium-scale system
  [panel (b)], the distributions shown in red indicate that simulated
annealing did not sample high-energy crystal structures, but selectively
sampled energetically stable structures. With no more than $M=30$
samples for the small-scale system and $M=250$ samples for the
medium-scale system, all crystal structures belonging to the
lowest-energy level were captured.
For the medium-scale system, the time required to obtain $M=250$ samples
was 0.0344$\left(3\right)$~s. In contrast, 0.646~s was required to
compare the Ewald energies by exhaustive search, corresponding
to the black histogram in the figure. Therefore, simulated annealing
enabled the stable structures to be searched approximately 18.8 times
faster.

\subsection{Quantum annealing for small- and medium-scale systems}
\label{sec:results.qa}

Having shown that annealing methods are effective for improving the
efficiency of the search, it is natural to examine whether quantum
annealing can achieve a more substantial speed-up. Annealing methods are
designed to enable global search by escaping from local minima through
mechanisms that overcome energy barriers~\cite{1983_SK_MV, 1998_TK_HN}.
In simulated annealing, transitions to higher-energy states are accepted
according to probabilities determined by the Boltzmann distribution,
allowing the system to overcome energy barriers within a classical
algorithmic framework~\cite{1983_SK_MV}. In contrast, quantum annealing
utilizes quantum superposition to explore many candidate states and
uses quantum tunneling to pass through energy barriers, thereby driving
the system toward the global optimum~\cite{1998_TK_HN}.
Because this procedure is expected to exploit quantum-mechanical
parallelism, it has been regarded as a method that can potentially
achieve a substantial speed-up over classical implementations
~\cite{2025Quinton_Zhang}.
However, in practical quantum annealing hardware, there are technical
limitations on both the number of physical qubits and the number of
couplings among them. Therefore, as the system size increases, the
speed-up achievable with current hardware is expected to become limited
~\cite{2025Quinton_Zhang}.

\vspace{2mm}
First, for the small-scale system CaYAlO$_4$, quantum annealing
selectively sampled low-energy structures, as shown in
Supplementary Fig.~S1. Since the search was performed with an annealing
time of 20~$\mu$s per sample, obtaining $M=30$ samples required
600~$\mu$s. The exhaustive search for this system was estimated to require
70{,}000~$\mu$s; thus, quantum annealing achieved a speed-up of 116 times
compared with exhaustive search.

\begin{figure*}[t]
  \centering
  \subfigure[20~$\mu$s]{%
    \includegraphics[width=0.3\linewidth]{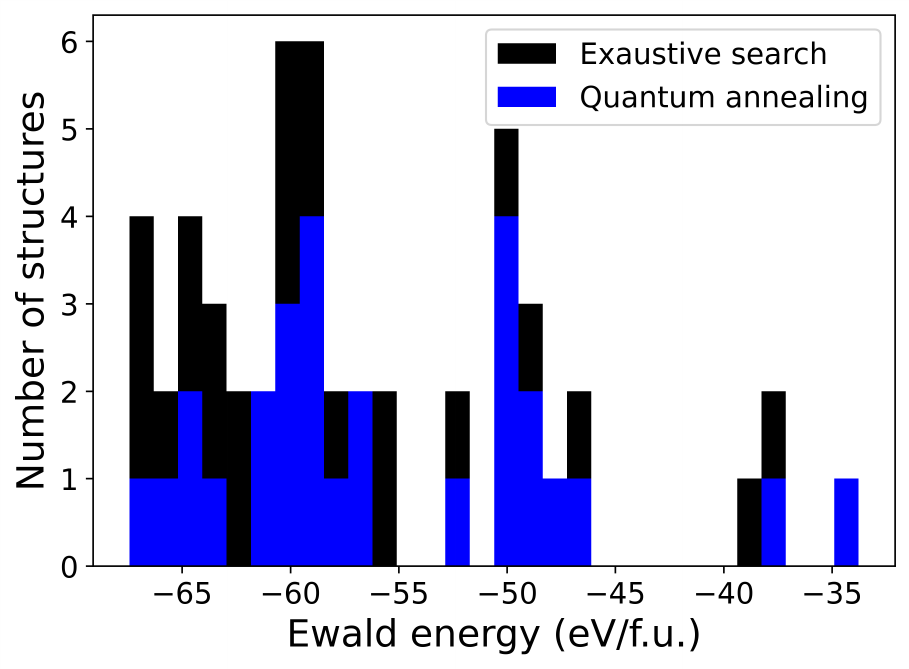}%
    \label{fig:KSbF4_20us_QA_unique_hist}}
  \hfill
  \subfigure[200~$\mu$s]{%
    \includegraphics[width=0.3\linewidth]{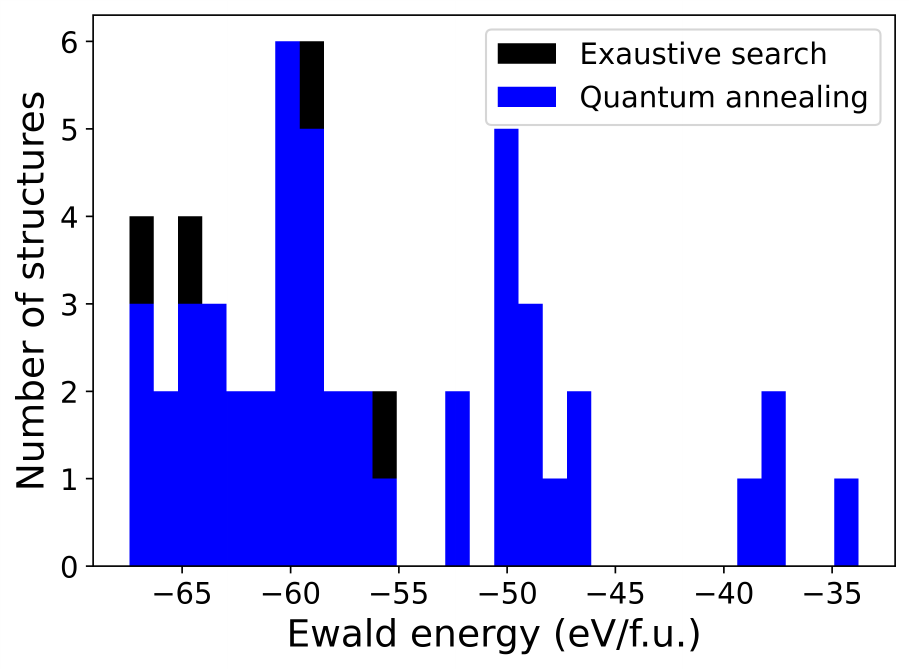}%
    \label{fig:KSbF4_200us_QA_unique_hist}}
  \hfill
  \subfigure[2{,}000~$\mu$s]{%
    \includegraphics[width=0.3\linewidth]{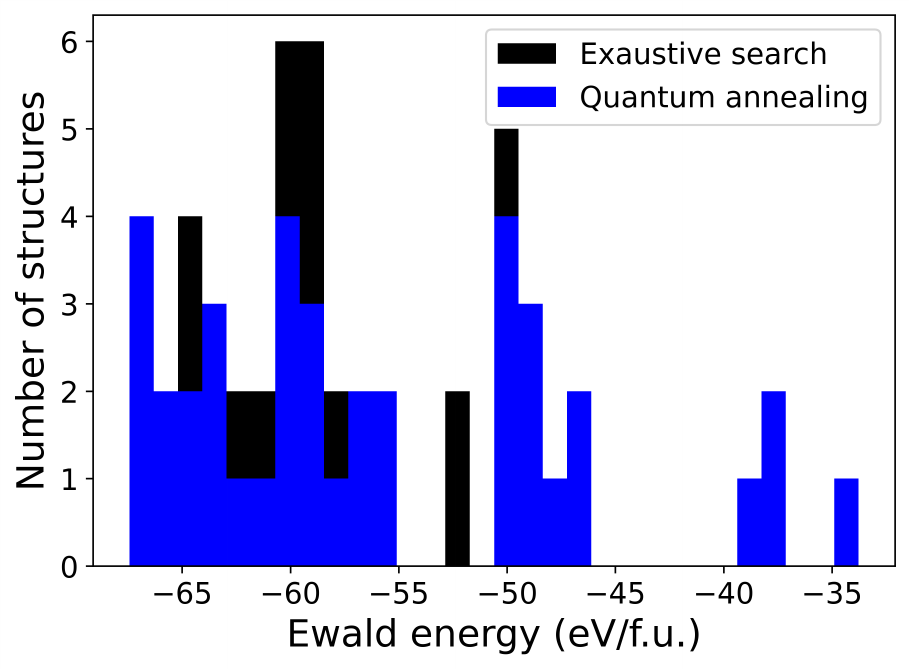}%
    \label{fig:KSbF4_2000us_QA_unique_hist}}
  \caption{
    Distribution of Ewald energies for symmetrically inequivalent
    structures in the medium-scale system $\beta$-KSbF$_4$ obtained by
    exhaustive search (black), and distributions of structures obtained by
    quantum annealing (blue) with annealing times of
    (a) 20~$\mu$s, (b) 200~$\mu$s, and (c) 2{,}000~$\mu$s.
    The number of samples was fixed at 250, and the Ewald energy is
    given in eV per formula unit.
  }
  \label{fig:QA_time_dependency_unique}
\end{figure*}

\vspace{2mm}
Next, the results obtained by applying quantum annealing to the
medium-scale system $\beta$-KSbF$_4$ are shown in
Fig.~\ref{fig:QA_time_dependency_unique}. Results are shown for three
annealing times: $\tau=20$~$\mu$s [panel (a)],
$\tau=200$~$\mu$s [panel (b)], and $\tau=2{,}000$~$\mu$s [panel (c)].
For $\tau=20$~$\mu$s [panel (a)], which showed sufficient search
performance for the small-scale system (Supplementary Fig.~S1), 
the crystal structures belonging to the lowest-energy range were not fully captured;
the blue histogram obtained by quantum annealing does not fully cover
the black histogram obtained by exhaustive search.
As $\tau$ was increased and the search time was extended,
the low-energy structures were captured more
completely. At $\tau=2{,}000$~$\mu$s [panel (c)], all crystal structures
in the lowest-energy range were captured.
In this case, the calculation required 0.5~s. Since exhaustive search is
known to require 0.646~s, the speed-up achieved by quantum annealing
under this condition is reduced to a factor of 1.29.

\subsection{Simulated annealing results and computational cost
  for the large-scale system}
\label{sec:results.large}
\begin{figure}[t]
  \centering
  \includegraphics[width=0.7\linewidth]
  {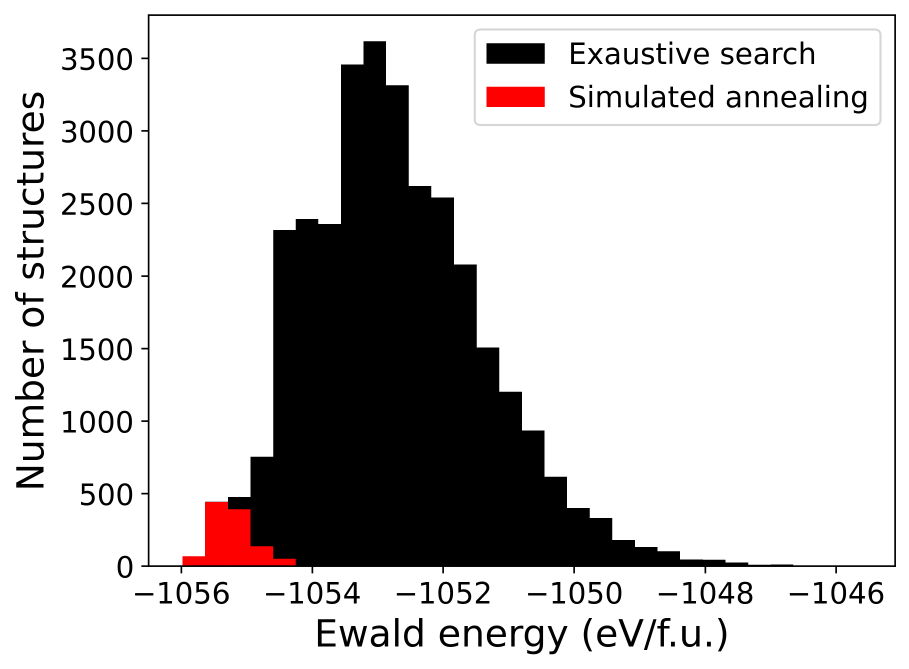}
  \caption{
    Distribution of Ewald energies for all symmetrically
    inequivalent mixed-crystal structures in the large-scale Ba-doped
    SiAlON system obtained by exhaustive search (black), and the
    distribution of structures obtained by simulated annealing (red).
    The Ewald energy is given in eV per formula unit.
    The red distribution fully covers the low-energy region, indicating that
    low-energy substitutional-site configurations were captured without
    omission.
  }
  \label{fig:Ba-SiAlON_SA_unique_hist}
\end{figure}

In terms of omission, the performance of quantum annealing was limited
at the size of the medium-scale system $\beta$-KSbF$_4$, whereas
simulated annealing showed good performance. We therefore examined
whether this performance could be maintained for a larger system.
As the target system, we selected the large-scale Ba-doped SiAlON
system with a $2\times 2 \times 1$ supercell. This system contains
24 substitutional sites, yielding 127{,}400 possible substitutional
configurations and 31{,}976 symmetrically inequivalent structures.
The number of samples was set to $M=5{,}000$. The results are shown in
Fig.~\ref{fig:Ba-SiAlON_SA_unique_hist}.

\vspace{2mm}
Because this system has a large total number of configurations, the
histogram bins are distributed almost continuously. In simulated
annealing, taking 5{,}000 samples was sufficient to capture all
lowest-energy substitutional structures without omission. This sampling
was completed in 1.010$\left(6\right)$~s. In contrast, exhaustive search
required 289.07~s in total. Thus, simulated annealing achieved sampling
that was 286 times faster.

\vspace{2mm}
The main focus of this paper is the speed-up achieved by annealing
methods. As shown in the results, speed-ups of more than two orders of
magnitude can be achieved when the performance is evaluated solely in
terms of computational time. Nevertheless, there is an additional human
cost associated with reformulating the original problem as a Hamiltonian
suitable for annealing.
As discussed in the Introduction, if the conventional exhaustive search
is itself infeasible, this human cost does not constitute a serious
drawback, because there is no practical alternative for overcoming the
combinatorial explosion. Even so, it is meaningful to clarify the human
cost required to construct the Hamiltonian for annealing.

\vspace{2mm}
When the CIF file of the parent crystal, the substituting ions, and the
substitution ratios are provided as input to the \texttt{Pymatgen}
library~\cite{2013Ong_Ceder}, the Ewald-energy term in
Eq.~\eqref{eq.ewald} can be constructed. In the case of Ba-doped SiAlON,
for which the number of possible substitutional configurations reaches
 127{,}400, this construction requires only 0.78~s.
The constraint term can also be constructed from the same information
using the library; this step requires 0.000097~s. By passing the
calculated objective and constraint terms to the Python library
\texttt{pyqubo}~\cite{2022Zaman_Tanaka}, the Ising Hamiltonian including
the constraint term can be constructed; this step requires 0.00087~s.
Simulated annealing is then performed for the resulting Ising
Hamiltonian using the \texttt{dwave-samplers} library.
All Python libraries used in this procedure are publicly available.
Thus, the overall workflow can be executed mechanically using automated
libraries and does not require substantial human effort.

\subsection{Comparison with Conventional Search Methods}
\vspace{2mm}
As noted above, the systems considered in the present study have at most
on the order of $10^5$ possible configurations and are therefore
sufficiently small compared with those investigated in previous studies.
For example, Tateyama et al. performed high-throughput evaluations of
electrostatic energies for systems with configuration spaces on the order
of $10^8\sim10^{74}$~\cite{2023_SJ_YT}.
They used exhaustive enumeration for small system sizes
and, otherwise, optimized individual samples using a greedy,
swap-based local optimization algorithm.
Using FeSbO$_4$ as an example, they confirmed that the minimum-energy 
structure obtained by exhaustive enumeration of the $3 \times 3 \times 2$ 
supercell was reproduced through a greedy search for the 
$3 \times 3 \times 14$ supercell. However, they did not verify whether 
the structure identified for the $3 \times 3 \times 14$ supercell
corresponded to the global minimum for that supercell size.

\subsection{Incomplete Sampling and Chain Breaks in Quantum Annealing}
\label{sec:results.omission}
Revisiting Table~\ref{tab:target_system}, quantum annealing, which was
expected to show an advantage over simulated annealing, did not exhibit
a speed-up for problems of realistic size. This is, of course, not
necessarily an intrinsic limitation of quantum annealing itself, but is
rather considered to arise from technical limitations in the current
hardware implementation.
In the present study, however, no problem such as failure of embedding
due to an insufficient number of qubits occurred. Nevertheless, the
quantum annealing calculations exhibited the problem of missing optimal
solutions. This is one of the findings clarified in the present study.

\vspace{2mm}
There is a previous study that applied quantum annealing to
electrostatic-energy evaluation using an approach similar to ours
~\cite{2024TB_MHE}. In the large-scale Ba-doped SiAlON system considered
in the present study, the number of possible configurations is on the
order of $10^5$. In contrast, that study treated LiCoO$_2$, which has a
much larger number of possible configurations, on the order of $10^{10}$.
They reported successful optimization by quantum annealing by adding a
chemical-potential term, which modifies the curvature around the minimum
of the objective function and thereby helps the calculation capture the
optimal solution. Because their study targeted a system with a large
number of configurations from the outset, they did not compare their
results with exhaustive search over all configurations, as performed in
the present study. Instead, they estimated the overall shape of the
energy distribution over the target configuration space using the
replica-exchange Monte Carlo method~\cite{1996Hukushima_Nemoto}.
The replica-exchange Monte Carlo method samples various solutions
according to Boltzmann distributions at different temperatures and
exchanges samples obtained at different temperatures with a certain
probability. This enables the search to escape from local optima and
explore low-energy solutions. Therefore, rather than being equivalent to
exhaustive search, this method is closer in spirit to the simulated
annealing used in the present study, in the sense that it examines the
distribution of accessible configurations by varying the temperature.
Consequently, the previous study cannot be taken as conclusive evidence
that low-energy substitutional configurations were exhaustively captured
by quantum annealing. In the present study, we have clearly demonstrated
that quantum annealing does not necessarily capture low-energy
substitutional configurations efficiently and without omission.

\vspace{2mm}
It is then important to identify what constitutes the bottleneck that
causes quantum annealing to fail to capture all low-energy
substitutional configurations. It is also an important question whether
such a problem can be overcome by future improvements in quantum
annealing hardware implementations.
A clear difference between the results for the small-scale system
CaYAlO$_4$ and the medium-scale system $\beta$-KSbF$_4$ is the presence
or absence of chain breaks. For the small-scale system CaYAlO$_4$, no
chain breaks occurred in any of the solutions. In contrast, for the
medium-scale system $\beta$-KSbF$_4$, chain breaks occurred in some of
the solutions. In this case, the chain break fraction was distributed
over the range from 5.55\% to 33\%. Benchmark studies have shown that
when chain breaks occur, it becomes more difficult to reach low-energy
solutions~\cite{2022Carugno_Cremonesi}. These observations suggest that
chain breaks may be one of the factors that made the optimization of
this system difficult.

\begin{figure}[t]
  \centering
  \includegraphics[width=0.7\linewidth]
  {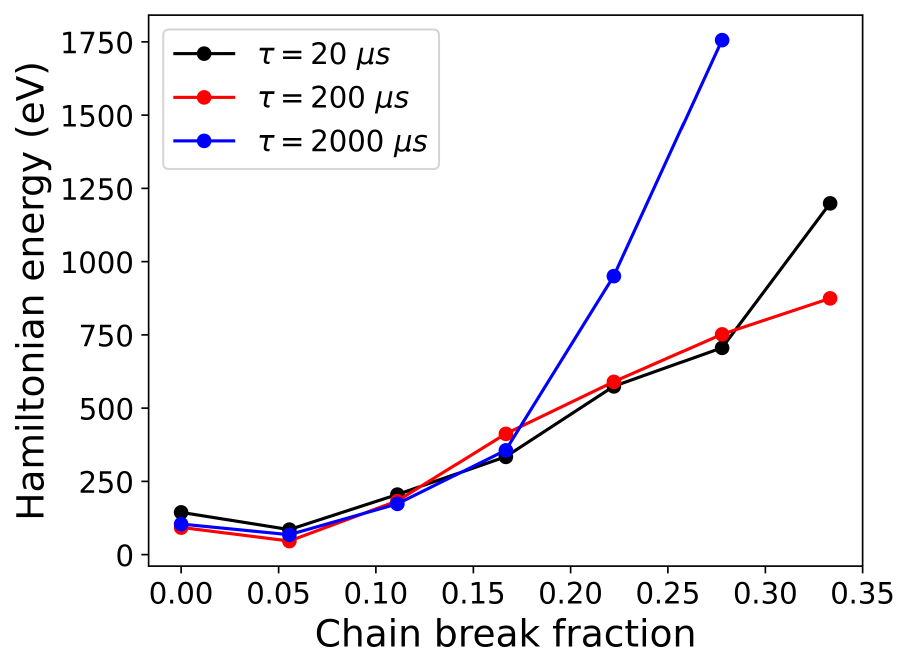}
  \caption{
    Relationship between the chain break fraction and
    solution quality for the medium-scale system $\beta$-KSbF$_4$.
    The horizontal axis represents the chain break fraction,
    and the vertical axis represents the Hamiltonian energy of
    the solutions, including the constraint term, 
    in units of eV. Because solutions that do not satisfy 
    the constraint term are also included, the energy is not 
    given in eV per formula unit.
  }
  \label{fig:betaKSbF4_chainbreak}
\end{figure}

\vspace{2mm}
We therefore examined whether chain breaks hinder the search for
low-energy solutions in the medium-scale system $\beta$-KSbF$_4$ using
an analysis method similar to that used in the previous study. The
results are shown in Fig.~\ref{fig:betaKSbF4_chainbreak}. In this
analysis, the chain break fraction is plotted on the horizontal axis,
and the average energy of the sampled solutions at each chain break
fraction is plotted on the vertical axis, allowing their correlation to
be examined~\cite{2022Carugno_Cremonesi}.
As in the previous study, the results show an overall positive
correlation. This supports the interpretation that, as chain breaks
become more frequent, the search for low-energy solutions is hindered,
resulting in a higher average energy of the sampled solutions. Although
there are local regions where the average energy decreases, similar
behavior was also observed in the previous 
study~\cite{2022Carugno_Cremonesi}.

\vspace{2mm}
The reason why chain breaks become a serious problem can be explained as
follows. Because the electrostatic interaction is long-ranged, interactions
arise between all pairs of charges in the simulation cell. In the present
formulation, the charges are represented by binary variables, and therefore
interactions are defined between all pairs of binary variables in the
Hamiltonian. As the number of binary variables in the Hamiltonian increases,
both the number of qubits required to represent them and the number of
couplings between them increase.
On the other hand, because the connectivity between physical qubits
implemented in a quantum computer is limited, long chains of physical qubits
must be constructed to realize the required connectivity between binary
variables. Moreover, these chains must be embedded so that the required
interactions between binary variables can be realized, making the
embedding increasingly complex. Thus, unless a quantum computer
provides sufficiently dense qubit connectivity, the Hamiltonian considered
here requires long chains and complex minor embeddings~\cite{2015Vinci_Lidar}.
As a result, chain breaks become a major obstacle to obtaining solutions.
This explains why the efficiency of QA decreases rapidly relative to SA
as the system size increases,
even for the intermediate-sized systems considered in the present study.

\vspace{2mm}
Low-energy substitutional structures extracted by quantum annealing were
subject to omissions, and not all low-energy substitutional structures
could be covered. Nevertheless, in practical applications, the demand is
often not to ensure that the true optimal structure with an even lower
energy than a given low-energy substitutional structure is captured, but
rather to generate plausible low-energy substitutional structures, even
partially, and proceed first to property evaluation.
This pragmatic approach, in which promising candidate structures are
selected as needed rather than exhaustively covering an enormous number
of possibilities, is conceptually similar to the special quasirandom
structures (SQS)~\cite{1990Zunger_Bernard}. If the missed substitutional 
patterns are of concern, it would also be possible to apply another 
heuristic method, such as a genetic algorithm using the same objective 
function, in a subsequent stage.

\section{Conclusion}\label{sec.conclusion}
In this study, we systematically examined the effectiveness of applying
annealing methods to the pre-screening of a vast substitutional
configuration space in solid solutions and disordered systems, using
electrostatic energy, specifically Ewald energy, as the objective
function. Specifically, the occupation states of substitutional sites
were represented by 0/1 binary variables, and the Ewald energy was mapped
onto an Ising-type Hamiltonian. This formulation allowed the search for
low-Ewald-energy configurations to be treated as a combinatorial
optimization problem.
We implemented this formulation using simulated annealing (SA) and
quantum annealing (QA). Through comparison with exhaustive search, we
quantitatively evaluated the applicability of annealing methods in terms
of the degree of search acceleration and whether substitutional
structures giving the lowest energy could be captured without omission.

\vspace{2mm}
For the small-scale system CaYAlO$_4$, both SA and QA sampled almost no
high-energy configurations and were confirmed to selectively extract
structures in the low-energy region. For this system, SA captured the
lowest-energy bin without omission with a speed-up of nearly 30 times
relative to exhaustive search. QA also captured the lowest-energy bin
without omission, while achieving a speed-up of more than 100 times.
These results demonstrate that, when the problem size is sufficiently
small, QA can also function as a practical high-speed search method.

\vspace{2mm}
In contrast, for the medium-scale system $\beta$-KSbF$_4$, SA showed a
significant speed-up, although its performance depended on the cooling
schedule. In particular, linear cooling was found to achieve both a
speed-up on the order of 200 times and suppression of omissions in the
lowest-energy bin. 
Furthermore, even for the larger-scale Ba-doped SiAlON system, 
SA was able to capture the lowest-energy structures without omission 
with a speed-up of nearly 300 times. 
These results indicate that SA is an effective pre-screening method capable of
extracting low-energy configuration groups within realistic computational
resources, even for problems whose exhaustive evaluation is practically
infeasible.

\vspace{2mm}
In contrast, for QA applied to $\beta$-KSbF$_4$, the lowest-energy bin
could be captured by using a sufficiently long annealing time. However,
the speed-up was limited in terms of computational time, and omissions
occurred under short-annealing-time conditions. 
In this study, we found a positive correlation between the chain break
fraction and the average energy of the obtained solutions, consistent
with chain breaks hindering the search for low-energy solutions. 
These results suggest that, in current hardware implementations of QA, 
constraints associated with embedding can become a bottleneck 
in terms of both search performance and computational speed.

\vspace{2mm}
Overall, among the approaches examined here, SA showed comparatively
robust performance for Ewald-energy-based configuration search in
mixed crystals.
It was shown to achieve substantial improvements in
computational efficiency even for large-scale systems while suppressing
omissions. In contrast, although QA is promising for small-scale
problems, improving implementation and embedding techniques, including
the reduction of chain breaks, is an important issue for achieving stable
performance in problems of realistic size.

\vspace{2mm}
The formulation used in this study can be implemented mechanically using
publicly available libraries and is practical for incorporating rapid
pre-screening based on Ewald energy into materials-design pipelines. By
using the low-energy structures extracted from Ewald-energy
screening as starting points and integrating them stepwise with
first-principles calculations or other search methods, this framework can
contribute to further acceleration of candidate-structure generation and
property prediction for real materials systems.

\section*{Acknowledgments}
This work was supported by JST SPRING, Japan (Grant Number JPMJSP2102), 
and JST BOOST, Japan (Grant Number JPMJBS2425).
The computations in this work have been performed using the facilities
of Center for Advanced Scientific Computing at JAIST.
K.H. is grateful for financial support from MEXT-KAKENHI,
Japan (JP23K23438, JP24K07571, and JP25K01851) and
JSPS Program for Forming Japan's Peak Research Universities
(J-PEAKS; JPJS00420230006).
R.M. is grateful for financial support from
MEXT-KAKENHI (22H051462, 24K01172, 24K07571A).
T.I. appreciates the support from the JSPS KAKENHI
Grant Number 24K17618 and JSPS Overseas Research Fellowships.

\section*{Data Availability}
The data supporting the findings of this study are available 
from the corresponding author on reasonable request.

\section*{Code availability}
The custom code used to generate the results of this study is available
from the corresponding author on reasonable request.

\section*{Use of AI-assisted tools}
AI-assisted tools were used during the preparation of this manuscript
for coding assistance, language editing, and translation.
All outputs were reviewed and verified by the authors, who take full
responsibility for the content of the manuscript.

\section*{Author contributions}
T.S. performed the investigation, curated and visualized the data, 
and wrote the original manuscript. K.H. contributed to the supervision 
of the study, provided resources, acquired funding, and reviewed and 
edited the manuscript. R.M. contributed to the supervision of the study, 
provided resources, and reviewed and edited the manuscript. 
T.I. supervised and administered the project and reviewed and 
edited the manuscript. All authors reviewed and approved 
the final manuscript.

\section*{Additional Information}
\subsection*{Competing interests}
The authors declare no competing interests.

\Urlmuskip=0mu plus 1mu
\bibliography{references}

\end{document}



\begin{center}

{\Large\bfseries Supplementary Information}

\vspace{1.5em}

{\large\bfseries
Application and Performance Assessment of Annealing Methods for
Electrostatic-Energy-Based Configuration Search in Mixed Crystals
}

\vspace{1.5em}

Tack Saquai$^{1}$,
Kenta Hongo$^{2}$,
Ryo Maezono$^{3}$,
and Tom Ichibha$^{1}$

\vspace{1em}

{\small
$^{1}$School of Information Science, JAIST,
Asahidai 1-1, Nomi, Ishikawa 923-1292, Japan\\
$^{2}$Center for Advanced Scientific Computing, JAIST,
Asahidai 1-1, Nomi, Ishikawa 923-1292, Japan\\
$^{3}$Graduate Major in Materials and Information Sciences,
School of Materials and Chemical Technology,
Institute of Science Tokyo,
2-12-1-S6-22 Ookayama, Meguro-ku, Tokyo 152-8550, Japan
}

\end{center}

\vspace{2em}

\subsection*{Computational environment}
\begin{table}[H]
  \centering
  \caption{
    Hardware specifications and major software versions used in this study.
    The hardware specifications refer to the computer used for exhaustive 
    search and simulated annealing.
  }
  \label{tab:pc_spec}
  \begin{tabular}{ll}
    \toprule
    Item & Specification \\
    \midrule
    \multicolumn{2}{l}{\textit{Hardware}} \\
    CPU                 & Apple M3 \\
    Number of CPU cores & 8 \\
    Memory              & 16 GB \\
    \addlinespace
    \multicolumn{2}{l}{\textit{Software}} \\
    Python              & 3.9.19 \\
    pymatgen            & 2024.8.9 \\
    PyQUBO              & 1.4.0 \\
    dwave-samplers      & 1.3.0 \\
    D-Wave Ocean SDK    & 8.0.1 \\
    dwave-system        & 1.26.0 \\
    \bottomrule
  \end{tabular}
\end{table}

\subsection*{Quantum annealing for the small-scale system}
\begin{figure}[H]
  \centering
  \includegraphics[width=0.65\linewidth]
    {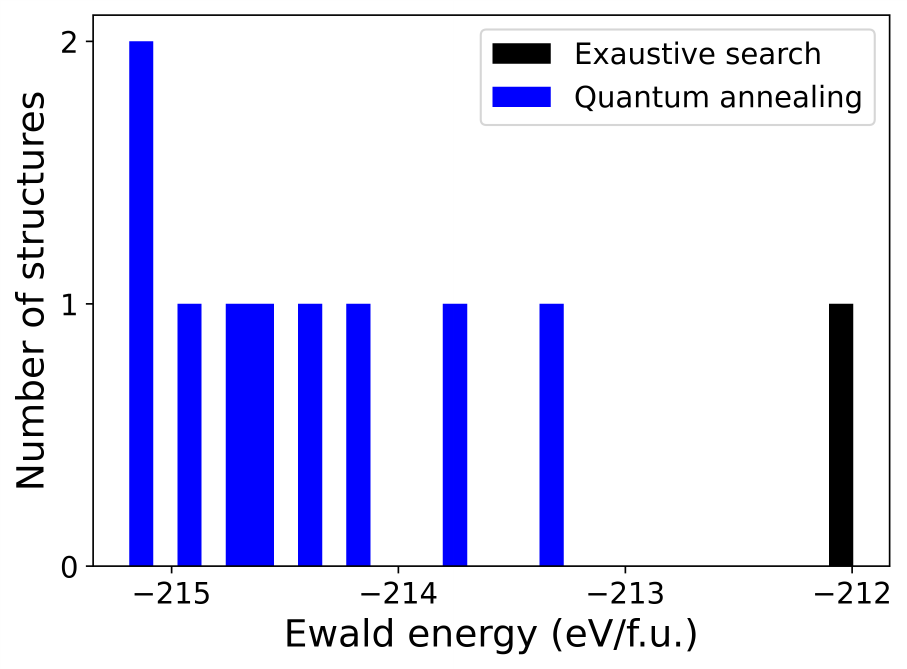}
  \caption{
    Distribution of Ewald energies for symmetrically inequivalent
    structures in the small-scale system CaYAlO$_4$ obtained by
    exhaustive search (black) and quantum annealing (blue).
    The Ewald energy is given in eV per formula unit.
  }
  \label{fig:CaYAlO4_QA}
\end{figure}
\section{Determination of $A_{ij}$, $B_{kl}$, and $C_{ki}$}
We briefly describe how to evaluate the coefficients,
$A_{ij}$, $B_{kl}$, and $C_{ki}$
of Eq.~(1) for the given disordered crystal structure.

\vspace{2mm}
Because the electrostatic energy can be expressed as a sum of products
of charges $q^{\prime}_{i}$, it can be written in the form
\begin{equation*}
  E = \sum_{j}^{}\sum_{i\leq j}^{}K_{ij}q^{\prime}_{i}q^{\prime}_{j}.
\end{equation*}
The coefficient $K_{ij}$ is a constant determined by the crystal
structure, including the relative positions of the atomic sites.
For example, \texttt{Pymatgen}~\cite{2013Ong_Ceder} supports the calculation of $K_{ij}$
from a crystal structure file and the subsequent evaluation of the
electrostatic energy from the charges assigned to the individual atomic
sites. The above expression also includes terms with $i=j$. There are two
reasons for including these terms. First, in a periodic system, an
electrostatic interaction arises between the charge $q^{\prime}_{i}$ at
atomic site $i$ and its periodic images. Second, the 
self-interaction correction must be taken into account
~\cite{1996Toukmaji_Board, 2022Shi_Shi}.
In the present study, the Ewald energy was used as the electrostatic energy. 
In the Ewald formulation, auxiliary charge distributions are introduced to 
achieve efficient convergence of the long-range interactions. 
The electrostatic energy arising from this procedure must be subtracted 
as a correction term. This correction is referred to as the self-interaction term 
and is proportional to $q_{i}^{\prime 2}$~\cite{1996Toukmaji_Board, 2022Shi_Shi}.

\vspace{2mm}
Here, the atomic sites can be classified into substitutional sites,
which undergo elemental substitution, and fully occupied sites without substitution.
Denoting the charges on the substitutional sites by $q_i$ and those on the fully occupied sites by
$Q_k$, the electrostatic energy is expressed as
\begin{equation*}
  E = \sum_{j}^{}\sum_{i\leq j}^{}A_{ij}q_{i}q_{j}+\sum_{k}^{}
  \sum_{l\leq k}^{}B_{kl}Q_{k}Q_{l}+\sum_{k}^{}\sum_{i}^{}C_{ki}Q_{k}q_{i}.
\end{equation*}
Here, the coefficients $A_{ij}$, $B_{kl}$, and $C_{ki}$ are constants
determined by the given crystal structure,
whereas $Q_k$ are constants determined by the elemental species occupying the site $k$.
The second term is constant, which equals the total electrostatic energy
among the charges on the fully occupied sites.
The third term represents the electrostatic energies
between substitutional sites $q_i$ and fully occupied sites $Q_k$.
The coefficient of $q_i$ is given by a constant $\sum_{k}C_{ki}Q_{k}$.

\vspace{2mm}
In the first component $\sum_{j}\sum_{i\leq j}^{}A_{ij}q_{i}q_{j}$,
the $i=j$ term, $A_{ii}q_{i}^2$, represents the electrostatic interactions
between $q_i$ and its periodic images, together with the self-interaction correction.
$A_{ii}$ is obtained as the electrostatic energy of a crystal in
which a unit charge is placed only at the atomic site $i$.
The $i \neq j$ term represents the electrostatic interactions
between $q_i$ and $q_j$, including their periodic images.
$A_{ij}$ is given by the electrostatic energy
of a crystal with unit charges placed only at atomic sites $i$ and $j$,
minus the self-energy contributions $A_{ii}$ and $A_{jj}$.

\vspace{2mm}
In the above procedure, auxiliary charge configurations that are not
electrically neutral are used to determine the coefficients~\cite{2022Shi_Shi}.
For these non-neutral configurations, no additional correction for the
net cell charge was included in the \texttt{Pymatgen} Ewald-energy
evaluation used in this study.

\section{Comparison between geometric and linear cooling in SA}
\label{sec.results.linear}

\begin{figure}[H]
  \centering
  \includegraphics[width=0.65\linewidth]
    {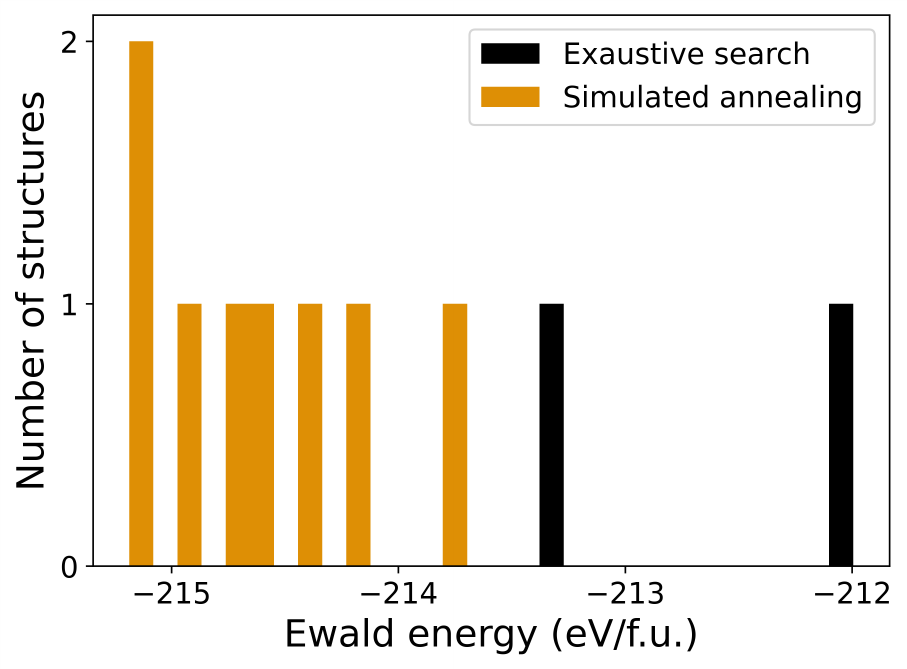}
  \caption{
    Distribution of Ewald energies for symmetrically inequivalent
    structures obtained by exhaustive search (black) and simulated
    annealing with linear cooling (orange) for
    CaYAlO$_4$. The Ewald energy is given in eV per formula unit.
  }
  \label{fig:SA_small_linear}
\end{figure}
\begin{figure}[H]
  \centering
  \includegraphics[width=0.65\linewidth]
    {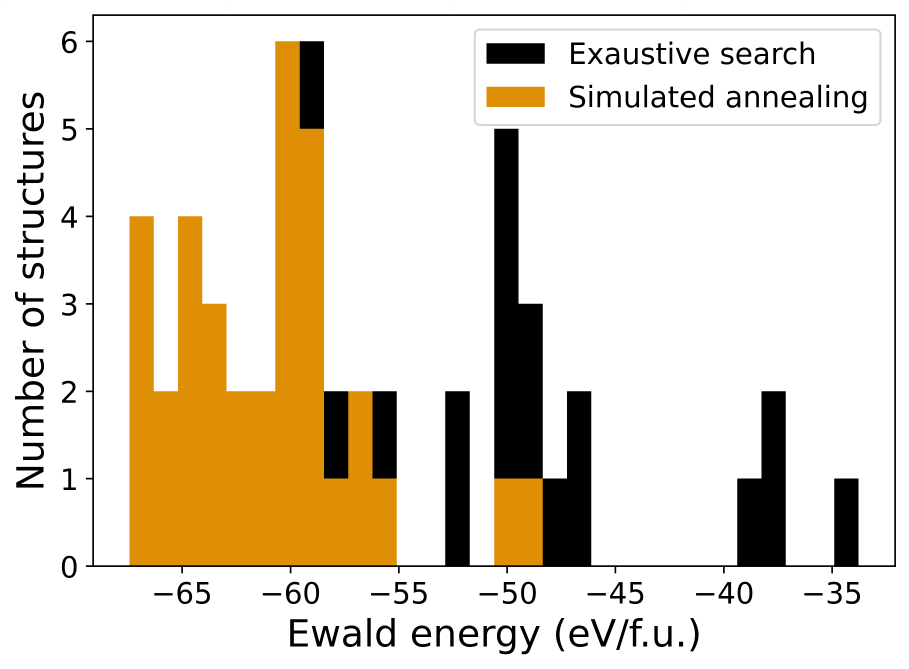}
  \caption{
    Distribution of Ewald energies for symmetrically inequivalent
    structures obtained by exhaustive search (black) and simulated
    annealing with linear cooling (orange) for
    $\beta$-KSbF$_4$. The Ewald energy is given in eV per formula unit.
  }
  \label{fig:SA_medium_linear}
\end{figure}
\begin{figure}[H]
  \centering
  \includegraphics[width=0.65\linewidth]
    {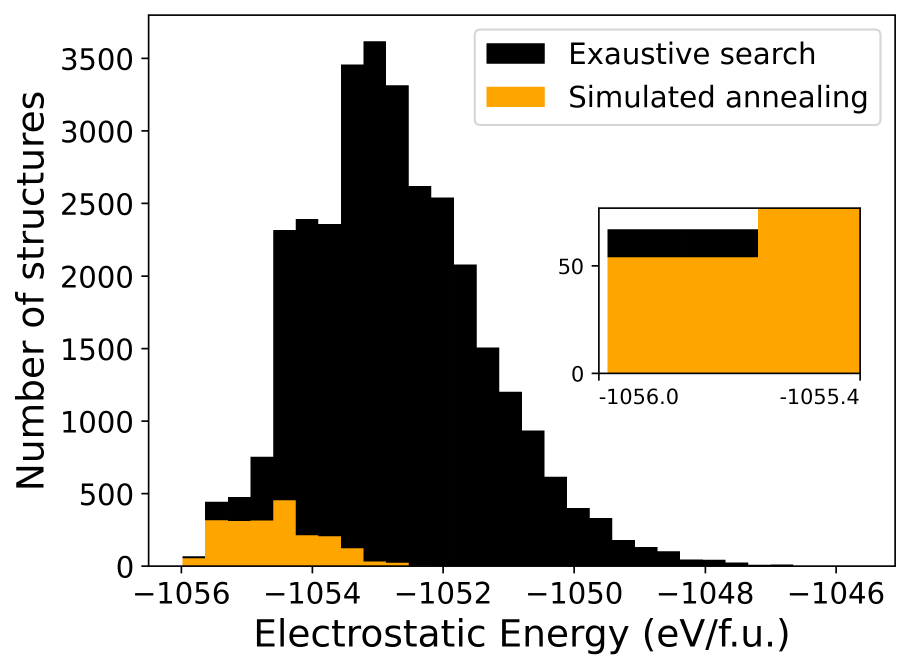}
  \caption{
    Distribution of Ewald energies for symmetrically inequivalent
    structures obtained by exhaustive search (black) and simulated
    annealing with linear cooling (orange) for
    Ba-doped SiAlON. The Ewald energy is given in eV per formula unit.
    The inset shows an enlarged view of the lowest-energy bin 
    for comparison between exhaustive search and simulated annealing.
  }
  \label{fig:SA_large_linear}
\end{figure}

The simulated annealing calculations presented so far
were performed using the default geometric cooling schedule.
When this schedule was replaced by linear cooling,
an improvement in computational speed was observed.
However, for the large-scale system,
it was also found that some low-energy structures were missed.

\vspace{2mm}
Supplementary Fig.~S2 shows the results obtained
by applying simulated annealing with linear cooling to the small-scale
system CaYAlO$_4$. The histogram indicates that the sampling was
concentrated in the low-energy region also for this system. In this case,
no more than $M=30$ samples were taken, and all structures belonging to
the lowest-energy bin were obtained. This sampling was completed in
0.00160$\left(2\right)$~s, corresponding to a speed-up factor of 43.8
relative to the 0.070~s required for exhaustive search.

\vspace{2mm}
Supplementary Fig.~S3 shows the results obtained
by applying simulated annealing with linear cooling to the medium-scale
system $\beta$-KSbF$_4$. The histogram indicates that the sampling was
concentrated in the low-energy region also for this system. In this case,
no more than $M=250$ samples were required to obtain all structures
belonging to the lowest-energy bin. This sampling was completed in
0.00260$\left(3\right)$~s, corresponding to a speed-up factor of 248
relative to the 0.646~s required for exhaustive search.

\vspace{2mm}
Supplementary Fig.~S4 shows the results
obtained by applying simulated annealing with linear cooling to the
large-scale Ba-doped SiAlON system. The histogram indicates that the
sampling was concentrated in the low-energy region also for this system.
In this case, using $M=5{,}000$ samples was not sufficient to obtain all
structures belonging to the lowest-energy bin, and some structures were
confirmed to be missed. This sampling was completed in
0.051$\left(3\right)$~s, which is 5{,}658 times faster than the
289.07~s required for exhaustive search. However, because some structures
were missed, this achieved performance is not listed in
Table~II of the main text.

\vspace{2mm}
As described above, linear cooling, similarly to geometric cooling,
sampled structures preferentially in the low-energy region. Except for
the large-scale system, it was able to obtain all structures belonging
to the lowest-energy bin without omission. It was also shown that linear
cooling can be evaluated faster than geometric cooling.

\vspace{2mm}
The main causes are considered to be the increased number of accepted
spin changes that raise the energy and the associated increase in the
time required to evaluate energy changes. In simulated annealing,
spin changes that increase the energy are also accepted with a
probability determined by the Boltzmann distribution.
For each spin change, it is therefore necessary to first evaluate
how much the energy would increase and then
calculate the corresponding Boltzmann factor.
In the \texttt{dwave-samplers} library, the energy change associated
with flipping each spin is stored in advance and used for the acceptance
test. Because this energy change depends on the spin configuration, when
a spin flip is accepted and the configuration changes, the stored energy
changes for the other affected spins must also be updated. In other
words, the energy changes are not recalculated from scratch at every
step; instead, the values for the relevant spins are sequentially updated
according to the accepted spin flips. Therefore, as the number of accepted
spin flips increases, the number of such update operations also increases,
resulting in an increase in the time required to update the energy
changes.

\vspace{2mm}
Compared with linear cooling, geometric cooling resulted in a larger
number of accepted spin flips that increased the energy, by factors of
2.60 for the small-scale system CaYAlO$_4$, 117.7 for the medium-scale
system $\beta$-KSbF$_4$, and 183.8 for the large-scale Ba-doped SiAlON
system, on average.
Correspondingly, the time required to update the energy changes was also
longer for geometric cooling, by factors of 2.59 for CaYAlO$_4$, 117.2
for $\beta$-KSbF$_4$, and 187.6 for Ba-doped SiAlON, on average.

\vspace{2mm}
The reason why energy-increasing changes occurred more frequently with
geometric cooling than with linear cooling can be understood as follows.
In the \texttt{dwave-samplers} library, the inverse temperature is
updated according to Eqs.~(5) and (6).
The inverse temperature gradually increases with the temperature step.
In linear cooling, the increment of the inverse temperature is constant
with respect to the temperature step. In contrast, in geometric cooling,
the increment is small in the early stage and becomes large in the final
stage.
In other words, geometric cooling keeps the system in the low-inverse-
temperature region for a longer period. In this region, the Boltzmann
factor corresponding to the inverse temperature becomes larger, and
energy-increasing spin flips are therefore more likely to be accepted.
Thus, because geometric cooling remains longer in the
low-inverse-temperature region than linear cooling,
a larger number of energy-increasing spin flips are
considered to have been accepted. As a result, the time required to update
the energy changes also increased, leading to a longer computational time
than that for linear cooling.

\vspace{2mm}
In addition, we confirmed that the omission observed for the large-scale
Ba-doped SiAlON system can be improved by the following two approaches.
The first approach is to increase the number of sweeps by a factor of
twenty to $N=20{,}000$. This corresponds to slower cooling. Under this
condition, $M=5{,}000$ samples were obtained for the large-scale system
in 0.90$\left(7\right)$~s, corresponding to a speed-up factor of 321
relative to exhaustive search, and no omission was observed.
Considering that geometric cooling achieved a speed-up factor of 286,
increasing the number of sweeps in linear cooling can be regarded as
more effective for the large-scale Ba-doped SiAlON system. On the other
hand, when the same condition was applied to the other systems, the
speed-up factors relative to exhaustive search were 2.26 for the
small-scale system CaYAlO$_4$ and 13.4 for the medium-scale system
$\beta$-KSbF$_4$. Although no omission was observed in either case, the
corresponding speed-up factors obtained with geometric cooling were 26.6
and 18.8, respectively. Therefore, under this condition, linear cooling
is slower than geometric cooling for these systems.

\vspace{2mm}
The second approach is to increase the number of samples. When the
number of samples was increased to $M=40{,}000$, no omission was observed,
and the search was completed in 0.433$\left(5\right)$~s.
This corresponds to a speed-up factor of 668 relative to
exhaustive search, which is faster than the speed-up factor of
286 obtained with geometric cooling.
For the large-scale Ba-doped SiAlON system, omission was eliminated by
using eight times more samples than in the case of geometric cooling.
Therefore, we also performed sampling for the small- and medium-scale
systems using the same eightfold increase in the number of samples. As
expected, no omission was observed, as in the original sampling. The
speed-up factors relative to exhaustive search were 5.69 for the
small-scale system CaYAlO$_4$ and 30.2 for the medium-scale system
$\beta$-KSbF$_4$.
Under this condition, linear cooling was faster than geometric cooling
for the medium-scale and larger systems, whereas it was slower for the
small-scale system.

\begin{table*}[htbp]
  \centering
  \caption{
    Results of simulated annealing with linear cooling.
    The values shown for these methods represent the speed-up factors
    achieved relative to exhaustive search. The speed-up naturally depends
    on the computational conditions used for the annealing method; therefore,
    the values reported here should be regarded as representative speed-up
    factors achieved in the present calculations.
    For comparison, the results of simulated annealing with geometric cooling
    (SA[G]) and quantum annealing (QA),
    listed in Table~II of the main text,
    are also shown. In the results for linear cooling and quantum annealing
    shown here, no omission of the lowest-energy group of structures was
    observed.
  }
  \label{tab:SA_Linear}
  \setlength{\tabcolsep}{3pt}
  \begin{tabular}{lcccc}
    \hline
    System                                                 &
    \shortstack{Geometric\\cooling (SA[G])} &
    \shortstack{Linear cooling\\with 20 times\\more sweeps} &
    \shortstack{Linear cooling\\with 8 times\\more samples} &
    \shortstack{Quantum\\annealing (QA)} \\
    \hline
    CaYAlO$_4$\\($1\times 2\times 2$)
                                                           & 26.6 & 2.26 & 5.69 & 116  \\
    $\beta$-KSbF$_4$\\($1\times 1\times 2$)
                                                           & 18.8 & 13.4 & 30.2 & 1.29 \\
    Ba-doped SiAlON\\($2\times 2 \times 1$)
                                                           & 286  & 321  & 668  & --   \\
    \hline
  \end{tabular}
\end{table*}

\begin{table*}[htbp]
  \centering
  \caption{
    Probability $\left(\%\right)$ of capturing the most stable
    energy bin without omission in simulated annealing and quantum annealing.
    The computational conditions for quantum annealing follow those under
    which no omission was confirmed in \S III~B of the main text.
  }
  \label{tab:annealing_success_rate}
  \begin{tabular}{lcccc}
    \hline
    System                                                 &
    \shortstack{Geometric\\cooling (SA[G])} &
    \shortstack{Linear cooling\\with 20 times\\more sweeps} &
    \shortstack{Linear cooling\\with 8 times\\more samples} &
    \shortstack{Quantum\\annealing (QA)} \\
    \hline
    CaYAlO$_4$\\($1\times 2\times 2$)
                                                           & 71.1 & 72.5 & 100  & 50.5 \\
    $\beta$-KSbF$_4$\\($1\times 1\times 2$)
                                                           & 99.2 & 98.5 & 100  & 64.7 \\
    Ba-doped SiAlON\\($2\times 2 \times 1$)
                                                           & 82.3 & 71.3 & 64.3 & --   \\
    \hline
  \end{tabular}
\end{table*}

\vspace{2mm}
The results of the comparison are summarized in
Supplementary Table~\ref{tab:SA_Linear}. As shown in the table, whether linear cooling 
or geometric cooling is faster depends on the size of the target system. 
Therefore, the superiority of one over the other cannot be determined 
uniquely on the basis of computational time alone. It is then meaningful 
to compare the two cooling schedules from the viewpoint of search performance, 
namely the probability of obtaining all structures belonging to the 
most stable energy bin without omission when the annealing method is applied
multiple times.
Here, multiple trials do not mean the $M$ combinatorial optimizations
corresponding to the number of samples. Rather, they mean repeating the
entire set of $M$ combinatorial optimizations multiple times. In other
words, for each annealing condition described above, we evaluated the
probability with which all structures belonging to the most stable
energy bin can be obtained without omission.

\vspace{2mm}
These results are summarized in Supplementary Table~\ref{tab:annealing_success_rate}.
For comparison, the results for simulated annealing with geometric
cooling and quantum annealing are also included.

\vspace{2mm}
For simulated annealing, linear cooling with an increased number of
sweeps outperformed geometric cooling only for the small-scale system
CaYAlO$_4$, whereas it performed worse than geometric cooling for the
other systems. Linear cooling with an increased number of samples
obtained the lowest-energy bin without omission for the small-scale
system CaYAlO$_4$ and the medium-scale system $\beta$-KSbF$_4$.
However, for the large-scale Ba-doped SiAlON system, the probability of
obtaining the lowest-energy bin without omission was lower than under
the other conditions, indicating that omissions were more likely to
occur.

\vspace{2mm}
Thus, for simulated annealing, no single condition can be regarded as
consistently superior to the others from both viewpoints: the speed-up
factor and the probability of avoiding omissions.

\vspace{2mm}
On the other hand, the quantum annealing results in Supplementary
Table~\ref{tab:annealing_success_rate} show lower values than those for
simulated annealing for both the small-scale system CaYAlO$_4$ and the
medium-scale system $\beta$-KSbF$_4$. Therefore, quantum annealing can
also be regarded as inferior to simulated annealing in terms of the
probability of avoiding omissions.

\bibliographystyle{naturemag}
\bibliography{references}